\documentclass[preprint2]{aastex6}

\usepackage{amsmath} 

\begin{document}

\title{Relationship of EUV Irradiance Coronal Dimming Slope and Depth to Coronal Mass Ejection Speed and Mass}

\author{James Paul Mason\altaffilmark{1} and Thomas N. Woods\altaffilmark{1}}
\affil{Laboratory for Atmospheric and Space Physics, University of Colorado at Boulder \\
3665 Discovery Drive \\
Boulder, CO 80303, USA}

\author{David F. Webb\altaffilmark{2}}
\affil{Institute for Scientific Research, Boston College \\
Newton, MA 02458, USA}

\author{Barbara J. Thompson\altaffilmark{3}}
\affil{NASA Goddard Space Flight Center \\
Greenbelt, MD 20771, USA}

\author{Robin C. Colaninno\altaffilmark{4}}
\affil{Space Science Division, Naval Research Laboratory \\
Washington, DC 20009-1231, USA}

\and
\author{Angelos Vourlidas\altaffilmark{5}}
\affil{The Johns Hopkins University Applied Physics Laboratory \\
Laurel, MD 20732, USA}

\altaffiltext{1}{james.mason@lasp.colorado.edu}


\begin{abstract}

Extreme ultraviolet (EUV) coronal dimmings are often observed in response to solar eruptive events. These phenomena can be generated via several different physical processes. For space weather, the most important of these is the temporary void left behind by a coronal mass ejection (CME). Massive, fast CMEs tend to leave behind a darker void that also usually corresponds to minimum irradiance for the cooler coronal emissions. If the dimming is associated with a solar flare, as is often the case, the flare component of the irradiance light curve in the cooler coronal emission can be isolated and removed using simultaneous measurements of warmer coronal lines. We apply this technique to 37 dimming events identified during two separate two-week periods in 2011, plus an event on 2010 August 7 analyzed in a previous paper, to parameterize dimming in terms of depth and slope. We provide statistics on which combination of wavelengths worked best for the flare-removal method, describe the fitting methods applied to the dimming light curves, and compare the dimming parameters with corresponding CME parameters of mass and speed. The best linear relationships found are 
\begin{eqnarray}
    v_{CME}\ \Big[\frac{km}{s}\Big] & \approx 2.36 \times 10^6\ \Big[\frac{km}{\%}\Big] \times s_{dim}\ \Big[\frac{\%}{s}\Big] \nonumber \\ 
    m_{CME}\ [g] & \approx 2.59 \times 10^{15} \Big[\frac{g}{\%}\Big] \times \sqrt{d_{dim}}\ [\%]. \nonumber
\end{eqnarray}
These relationships could be used for space weather operations of estimating CME mass and speed using near-realtime irradiance dimming measurements.

\end{abstract}


\section{Introduction} 
\label{sec:intro}
Large regions of temporary dimming or darkening of preexisting solar coronal emission often accompany coronal mass ejections (CMEs) and may trace field lines opened during the CME. The plasma of the solar corona responds in a number of ways to an eruptive event. \citet{Mason2014} provide details about the physics behind coronal dimming and the observational effects to be considered during analysis. Therein, the case is made for two hypotheses: that the slope of deconvolved, extreme ultraviolet (EUV) dimming irradiance light curves should be directly proportional to CME speed, and similarly, that dimming depth should scale with CME mass. Dimming regions can be extensive, representing at least part of the ``base" of a CME and the mass and magnetic flux transported outward by it. In this paper, we use the methods of \citet{Mason2014} to isolate EUV irradiance dimming as observed by Solar Dynamics Observatory (SDO; \citealt{Pesnell2012}) Extreme Ultraviolet (EUV) Variability Experiment (EVE; \citealt{Woods2012}) due to mass loss and characterize its time series in terms of slope and magnitude. \citet{Mason2014} focused on a single event and thus a correlation between dimming and CME parameters could not be established. Here, we analyze 37 events, 17 of which are used to establish a relationship between dimming slope and depth to CME speed and mass, respectively. We also outline the physical derivation for these relationships and assess their consistency with the data.

Extensive surveys of EUV images containing coronal dimming events and their relation to CMEs have been performed by \citet{Reinard2008, Reinard2009}. For their sample of ~100 dimming events, \citet{Reinard2008} found mean lifetimes of 8 hours, with most disappearing within a day. \citet{Reinard2009} studied CMEs with and without associated dimmings, finding that those with dimmings tended to be faster and more energetic.  \citet{Bewsher2008} found a 55\% association rate of dimming events with CMEs, and conversely that 84\% of CME events exhibited dimming.

The timescale for dimming development is typically several minutes to an hour. This is much faster than the radiative cooling time, which implies that the cause of the decreased emission is more dependent on density decrease than temperature change \citep{Hudson1996}. Studies have demonstrated that dimming regions can be a good indicator of the apparent base of the white light CME \citep{Thompson2000, Harrison2003, Zhukov2004}. Thus, dimmings are usually interpreted as mass depletions due to the loss or rapid expansion of the overlying corona \citep{Hudson1998, Harrison2000, Zhukov2004}. Many landmark studies have established that dimmings can contribute a large fraction of the mass to a CME \citep{Harrison2000, Harrison2003, Zhukov2004, Aschwanden2009}. 

An Earth-directed CME's potential geoeffectiveness is typically characterized by four values: its velocity, mass, and the magnitude and duration of the southward component of the magnetic field (B$_z$) impacting Earth.  Typical CME forecasts provide a predicted Earth arrival time only, which chiefly depends on velocity. The current standard process for estimating velocity relies on sequential coronagraph images from the Solar and Heliospheric Observatory's (SOHO; \citealt{Domingo1995}) Large Angle Solar Coronagraph C2 and C3 (LASCO; \citealt{Brueckner1995}) and the Solar Terrestrial Relations Observatory's (STEREO; \citealt{Kaiser2007}) COR1 and COR2 coronagraphs \citep{Howard2008}. Analysis of coronagraph images to determine CME velocities and masses results in relatively large uncertainties of 30-50\% \citep{Vourlidas2000, Vourlidas2010a, Vourlidas2011a}. The velocity and mass measurements with the most uncertainty are for Earth-directed CMEs that are seen as halos in coronagraphs at or near Earth. For these CMEs, speed determination is significantly affected by projection on the plane-of-sky, and a large percentage of the mass can be hidden behind the instrument's occulter. Without observations of these CMEs from another viewpoint, such as STEREO, it is difficult to make an accurate measurement of the CME velocity and mass from the coronagraph observations. However, EUV dimmings associated with these CMEs are very well observed by instruments in Earth orbit. 

Standard plane-of-sky velocity estimates are made and cataloged by the Coordinated Data Analysis Workshops (CDAW) CME catalog \citep{Gopalswamy2009}, which use routinely produced SOHO/LASCO coronagraph images. The different views from SOHO/LASCO and STEREO/COR images can be used to better constrain the velocity, direction, and mass of CMEs (e.g., \citep{Colaninno2009}). 

Coronal dimming can also be studied with spatially-integrated (full-disk) irradiance measurements as demonstrated by \citet{Mason2014}. They showed that a solar flare's impulsive and gradual phase peaks can initially dominate the irradiance for dimming-sensitive lines (e.g., Fe IX 171 \AA, Fe XII 195 \AA). They developed a technique for removing these flare peaks that only requires an independent, simultaneous irradiance measurement from a dimming-insensitive, flare-sensitive line (e.g. Fe XV 284 \AA), and they demonstrated how to apply this ``correction method" to the solar eruptive event on 2010 August 7. They also reported the kinetic energy parameters (mass and speed) of the associated coronal mass ejection (CME) that could be related to the dimming results of depth and slope. This follow-on paper expands that prior work to several more events in order to study the relationship of the CME and coronal dimming parameters.

Here, we analyze 37 coronal dimming events during two separate two-week periods during 2011 and search for the relationship between dimming and CME speed and mass, plus the event from the \citet{Mason2014} paper, for a total initial sample of 38 events. Of the events studied, 17 could be parameterized in both dimming with SDO/EVE data and in CME velocity from SOHO/LASCO and STEREO/COR observations. 14 of the events yielded valid results in terms of dimming with SDO/EVE data and CME mass derived from the coronagraph observations. Section 2 describes the method for selecting this sample of events and explains why some events initially identified in SDO's Atmospheric Imaging Array (AIA; \citealt{Lemen2012}) could not be analyzed with SDO/EVE and/or STEREO/COR. Section 3 provides statistics on the flare-peak correction method detailed in \citet{Mason2014}, specifically which combinations of dimming and non-dimming lines provided the best correction for each of the events. Section 4 describes the fitting method applied to the corrected SDO/EVE light curves, including a discussion of uncertainties. Finally, Section 5 shows the correlations between the various combinations of coronal dimming and CME parameters, and conclusions about dimming and CME relationships are presented in Section 6.

\section{Event Selection} \label{sec:eventselection}

\begin{figure}
    \plotone{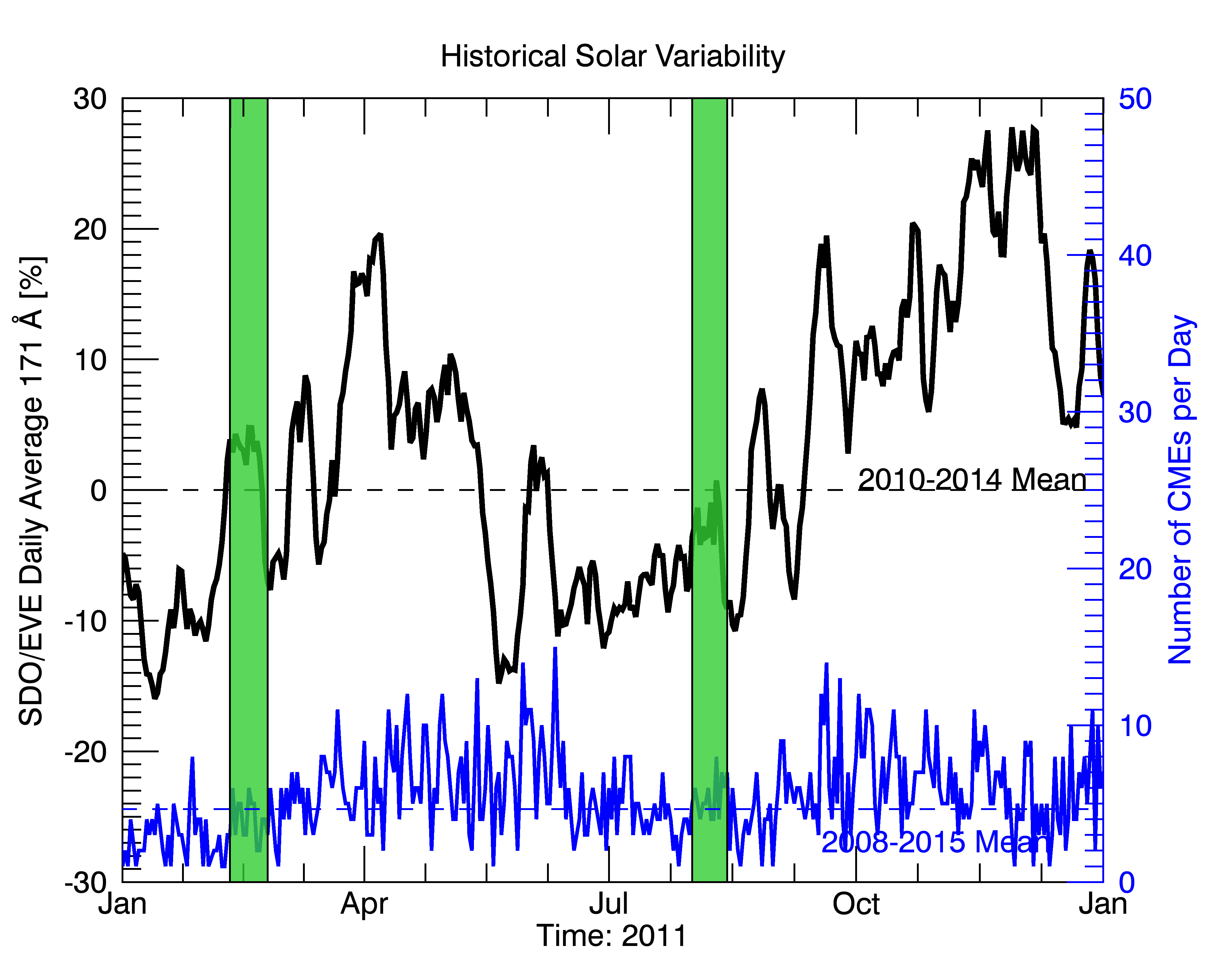}
    \caption{
        Context for the selected periods of study. The black line is the daily averaged EVE Fe IX 171 \AA\ line and the 
        blue line is the daily total CME occurrence. The vertical green bars indicate the selected periods of this study.
        The mean for EVE (dashed black line) is taken over the first four years of EVE's operations (2010-2014) and the 
        mean for CME occurrence (dashed blue line) is taken for the most recent solar cycle starting in 2008 to the end of 
        2015. Note that the full range of both of these means is not shown; only 2011 is shown for clarity of the selected
        periods. 
    }
    \label{fig:historicalcontext}
\end{figure}

Four weeks were selected in 2011 for analysis of coronal dimming events: February 10-24 and August 1-14 (Figure \ref{fig:historicalcontext}). These two independent periods about 6 months apart were chosen as appropriate times during the initial rise of solar activity during solar cycle 24. The initial criterion for this selection is to have a period of time that could result in more than 30 identifiable events. It is also desirable to select a time when the two STEREO spacecraft orbital locations were advantageous for geometric analysis, and when the other space-based instruments used in this study could be expected to be operating nominally. The periods of study are typical in terms of CME occurrence and solar EUV irradiance variability.

Images from SDO/AIA were used to first identify dimming events. Identification was performed manually using daily SDO/AIA movies to create a list of candidate events. The primary initial selection criteria were that 1) the dimming must persist for several hours and 2) the dimming have non-trivial spatial extent e.g., at least comparable to the size of an active region. The approximate time of the event was used to search the related observations in other instruments: flares from the Geostationary Operational Environmental Satellite (GOES) X-ray flux, CMEs from SOHO/LASCO and STEREO/COR, and solar irradiance from SDO/EVE. This initial list included 37 events. In some cases, the dimming was not clear in SDO/EVE data or the CME was not clearly identified in the coronagraph images; nevertheless these were dimmings identified in SDO/AIA and are listed in Table \ref{tab:eventlist} for completeness. Of these events, 29 could be parameterized with SDO/EVE, 21 had measured CME velocities, and 17 had measured CME masses. Six of the CMEs had at least two views so that 3-D analysis could be applied for improved accuracy of the CME kinetic parameters.

Because SDO/EVE irradiance observations are spatially integrated, dimmings from spatially-distant areas that occur too closely in time overlap in the irradiance time series and cannot be easily separated and parameterized. Thus, such events have a ``--" in the dimming columns of Table \ref{tab:eventlist} and are excluded from the correlative study in Section 5. This was the case for Events 9, 21, 29, 31, and 33. Secondly, some dimmings identified in SDO/AIA were not detectable in the SDO/EVE data making parameterization impossible. Here, ``not detectable" simply means that the EVE light curves did not show anything resembling the archetypal dimming near the time that was identified in AIA. This implies the magnitude of the dimming was small ($<$ 1\% impact on irradiance), which would be the case if the dimming itself was not very deep or if evolution elsewhere on the solar disk dominated (e.g., active region evolution). For example, Event 22 was a series of small eruptions from an active region with multiple slow CMEs whose analysis would be difficult; Event 24 was a very slight darkening of an active region's coronal loops with no identified CME; Event 25 was an off-disk dimming event with a narrow CME; and Event 28 was a small occurrence of ``coronal rain", also with no identified CME. In principle, it is possible for all of the event types described above to generate a large disk-integrated irradiance change, but in these cases the change was insufficient to be observable by SDO/EVE. In total, these criteria on SDO/EVE measurements resulted in 9 of the 38 events being excluded from the correlation analysis, leaving 29 events. Given the quality constraints on EVE dimming and coronagraph data, there were 17 events that were parameterized both in terms of dimming and CME kinetic parameters. 

\begin{deluxetable*}{ccccccccc}
\label{tab:eventlist}
\tablecaption{Event list. Times and locations are approximate. Only 29 of the events have dimming and CME derived parameters to allow the study of the relationships between dimmings and CMEs.}
\tablecolumns{9}
\tablehead{
\colhead{Event \#} &
\colhead{Date} &
\colhead{Time} & \colhead{Location} & \colhead{GOES} &
\colhead{Dimming} & \colhead{Dimming} &
\colhead{CME} & \colhead{CME} \\ 
\colhead{} & \colhead{} & \colhead{[UTC]} & \colhead{} & \colhead{Flare} & \colhead{$$Depth [\%]} & \colhead{Slope} & \colhead{Mass [g]} & \colhead{Speed} \\
\colhead{} & \colhead{} & \colhead{} & \colhead{} & \colhead{Class} & \colhead{} & \colhead{[\% s$^{-1}$]} & \colhead{} & \colhead{[km s$^{-1}$]}
}
\startdata
1 & 2011 Feb 10 & 07:40 & N20 W-limb & -- & 0.67 & 0.85 & -- & -- \\
2 & 2011 Feb 10 & 13:36 & N20 W-limb & -- & 0.21 & 1.52 & 3.40E+14 & 338 \\
3 & 2011 Feb 11 & 07:46 & N20 W-limb & B9.0 & 1.65 & 1.37 & 1.40E+14 & 175 \\
4 & 2011 Feb 11 & 13:21 & N60 W00 & -- & 0.81 & 0.51 & -- & -- \\
5 & 2011 Feb 11 & 21:43 & N10 E-limb & -- & 1.32 & 0.96 & 2.60E+15 & 469 \\
6 & 2011 Feb 12 & 06:05 & N30 E10 & -- & 1.05 & 3.26 & -- & -- \\
7 & 2011 Feb 13 & 14:00 & S10 E10 & M6.6 & 2.93 & 2.13 & 3.42E+15 & 349 \\
8 & 2011 Feb 14 & 15:45 & S10 W00 & C6.6 & 1.42 & 0.79 & 1.20E+13 & 303 \\
9 & 2011 Feb 14 & 17:36 & N30 E20 & M2.2 & -- & -- & 4.62E+15 & 396 \\
10 & 2011 Feb 15 & 02:07 & N00 W00 & X2.2 & 4.41 & 1.63 & 5.09E+15 & 897 \\
11 & 2011 Feb 16 & 14:40 & S20 W30 & M1.6 & 1.95 & 1.49 & -- & -- \\
12 & 2011 Feb 17 & 00:47 & E40 W00 & -- & 2.67 & 1.65 & -- & -- \\
13 & 2011 Feb 18 & 11:15 & S10 W50 & -- & 1.22 & 1.51 & 6.60E+13 & 350 \\
14 & 2011 Feb 18 & 19:20 & N30 W00 & C7.1 & 4.27 & 0.76 & -- & -- \\
15 & 2011 Feb 24 & 07:40 & N10 E-limb & M3.5 & 3.71 & 1.88 & -- & -- \\
16 & 2011 Feb 25 & 07:00 & N45 E60 & -- & 1.32 & 1.21 & 6.50E+14 & 370 \\
17 & 2011 Aug 2 & 05:10 & N05 W20 & M1.4 & 4.76 & 1.02 & 7.10E+15 & 1110 \\
18 & 2011 Aug 2 & 13:00 & N00 E-limb & -- & 0.47 & 2.22 & -- & -- \\
19 & 2011 Aug 3 & 13:43 & N05 W48 & M6.0 & 2.68 & 2.87 & 7.80E+15 & 1100 \\
20 & 2011 Aug 4 & 04:12 & N05 W58 & M9.3 & 5.22 & 3.54 & 5.70E+15 & 2080 \\
21 & 2011 Aug 4 & 04:41 & N80 W00 & -- & -- & -- & -- & 338 \\
22 & 2011 Aug 5 & 07:25 & S30 E50 & -- & -- & -- & -- & 110 \\
23 & 2011 Aug 6 & 11:50 & S14 E10 & C1.3 & 1.67 & 1.19 & -- & -- \\
24 & 2011 Aug 6 & 18:25 & N05 W25 & -- & -- & -- & -- & -- \\
25 & 2011 Aug 6 & 17:35 & N30 W-limb & C1.4 & -- & -- & 5.10E+14 & 176 \\
26 & 2011 Aug 6 & 22:40 & N10 W25 & -- & 0.99 & 0.43 & -- & -- \\
27 & 2011 Aug 7 & 04:00 & N10 W55 & -- & 0.6 & 1.13 & 7.00E+13 & 459 \\
28 & 2011 Aug 8 & 01:15 & N80 E05 & -- & -- & -- & -- & -- \\
29 & 2011 Aug 8 & 11:00 & N15 W70 & C1.3 & -- & -- & -- & -- \\
30 & 2011 Aug 8 & 17:42 & N05 W05 & -- & 2.6 & 2.68 & -- & -- \\
31 & 2011 Aug 8 & 18:42 & N05 W75 & M3.5 & -- & -- & 2.09E+15 & 1248 \\
32 & 2011 Aug 9 & 08:10 & N15 W70 & X6.9 & 1.72 & 3.5 & 3.84E+15 & 1474 \\
33 & 2011 Aug 9 & 09:12 & S30 E-limb & -- & -- & -- & -- & -- \\
34 & 2011 Aug 9 & 11:26 & N05 W00 & -- & 2.79 & 1.42 & -- & 428 \\
35 & 2011 Aug 11 & 10:23 & N00 W-limb & C6.2 & 1.07 & 1.88 & 2.29E+15 & 1144 \\
36 & 2011 Aug 12 & 00:09 & N45 E80 & -- & 1.91 & 0.82 & 5.10E+14 & 346 \\
37 & 2011 Aug 12 & 11:13 & N50 E70 & -- & 1.22 & 0.66 & -- & -- \\
38 & 2010 Aug 7 & 18:05 & N05E60 & M1.0 & 2.18 & 1.57 & 6.40E+15 & 850
\enddata
\end{deluxetable*}

\section{Flare-Dimming Deconvolution Method Statistics} 
\label{sec:deconvolutionstatistics}

\begin{figure*}
    \begin{center}
	    \plotone{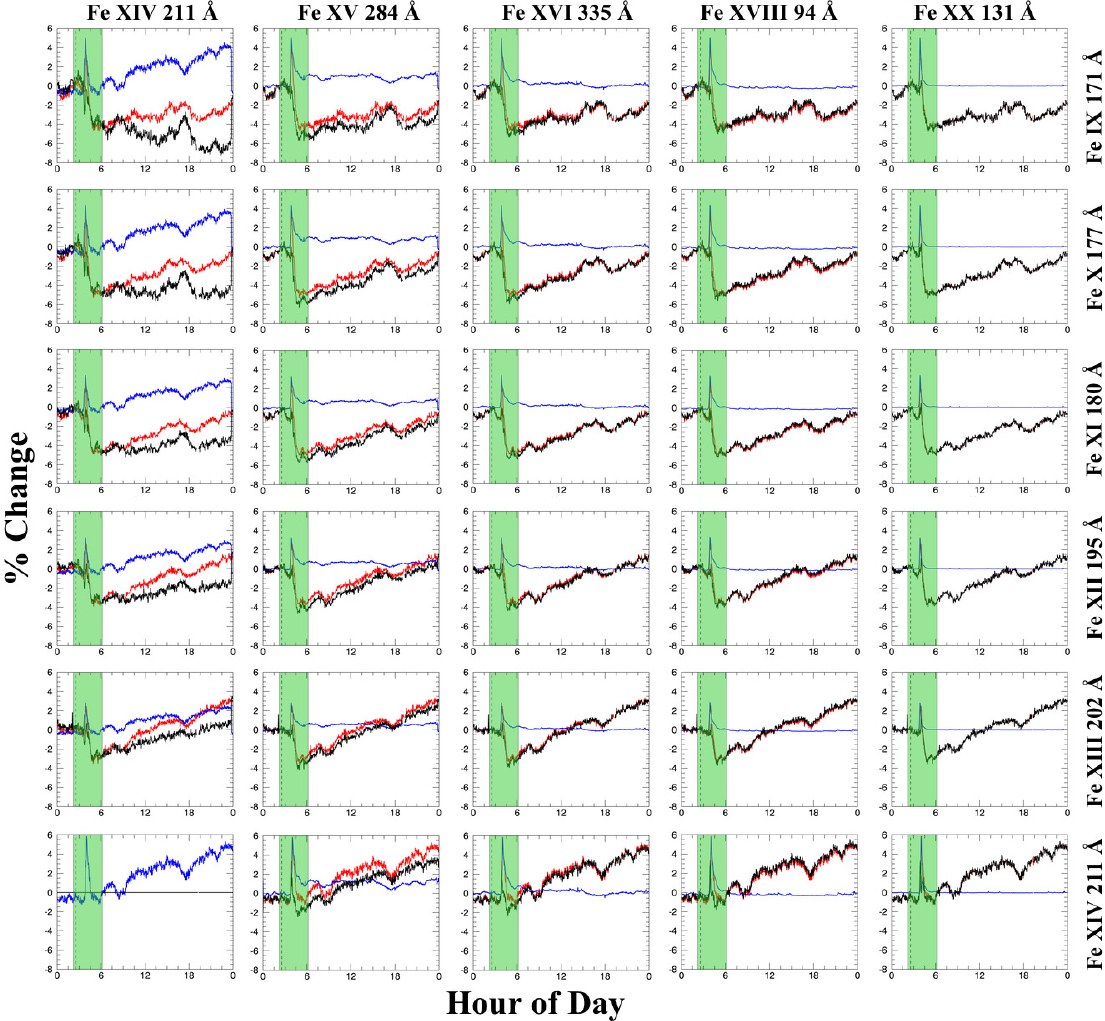}
    \end{center}
    \caption{
        Example of every combination of the dimming (rows) and non-dimming (columns) emission lines for the deconvolution 
        method for a single event (Event 20). In each plot, the red is the dimming line, blue is the scaled and time-shifted 
        non-dimming line, and black is the result of the subtraction (red - blue). The vertical transparent green bar 
        indicates the time window the algorithm uses for finding and matching peaks. All emission lines are for Fe. 
        Ionization state increases down for dimming lines and to the right for non-dimming lines. Each plot is over the same
        24-hour period. 
   	}
    \label{fig:deconvolutioncombinations}
\end{figure*}

There are 30 permutations of the dimming emission lines (171, 177, 180, 195, 202, 211 \AA) and non-dimming emission lines (211\footnote{Recall that 211 \AA\ is included in both dimming and non-dimming categories to reflect its ambiguity}, 284, 335, 94, 131 \AA) for the correction method. Each one is processed using the same algorithm described in \citet{Mason2014}. Figure \ref{fig:deconvolutioncombinations} shows an example of all 30 combinations for a single event (Event 20).

It can be seen that the higher the ionization state of the non-dimming line (blue), the ``purer" the flare light curve,  i.e., higher ionization states return almost perfectly back to their pre-flare irradiance level soon after the peak while lower ionization states show some additional post-peak response. Because the most intense heating occurs early in the flare -- during the impulsive phase as observed by GOES or the Reuven Ramaty High-Energy Solar Spectroscopic Imager (RHESSI; \citealt{Lin2002}) in hard x-rays -- it's unlikely that the emission from high ionization states disappears due to heating into the next ionization state. Rather, it returns to its pre-flare level because the intense heating supporting its existence is over and cooling has set in. Indeed, the mid-ionization states such as Fe XVI at 335 \AA\ show a slow, hours-long ramp downward in irradiance. The fact that these mid-ionization states don't immediately return back to their pre-flare level indicates that their net cooling rate is lower. The lower net cooling rate is likely due to the higher density of these ions where collisional de-excitation in the plasma competes with radiative cooling. Additionally, the Einstein A coefficient for Fe XVIII 94 \AA\ is 11.4x\footnote{Determined with the \href{http://physics.nist.gov/PhysRefData/ASD/lines_form.html}{NIST online database}} larger than for Fe XVI 335 \AA, indicating that the radiative cooling is even slower for Fe XVI 335 \AA. The blue light curve for FeXVI 335 \AA\ indicates that the cooling is ongoing during this hours-long period. In other words, warm ions like Fe XVI are slowly recombining with electrons and acting as a source term for the cooler ionization populations. Critically, this ``feeding" of the lower ionization populations, like Fe IX, is a cooling mechanism, not a mass-loss one. By removing this trend as indicated by the irradiance in e.g., Fe XV 284 \AA, we obtain a light curve more sensitive to mass-loss than temperature evolution (black curve in Figure \ref{fig:deconvolutioncombinations}).

In \citet{Mason2014}, it was found that for the 2010 August 7 event, the combination of Fe IX 171 \AA\ (dimming) and Fe XV 284 \AA\ (non-dimming) in EVE gave the best match to the spatially isolated dimming in AIA 171 \AA. The only dimming mechanisms identified to be important in this event were mass-loss and thermal. Thus, it seems that the 171 \AA\ - 284 \AA\ combination can successfully mitigate the impact of thermal processes on the dimming line. If other dimming mechanisms play an important role in the irradiance, it may be necessary to account for them, such as by identifying and removing the impact of obscuration dimming. Until such an analysis is performed, we apply the deconvolution method to the additional 28 events with viable EVE data, using the clean removal of the flare peak as the criteria for determining the best combination of dimming-non-dimming line. In other words, the peaks of the dimming and scaled/time-shifted non-dimming lightcurves should be similar in shape. Figure \ref{fig:deconvolutioncombinations} shows that many of the combinations would meet this criteria. The next determining factor is depth of dimming. Event 20 had a relatively consistent depth of dimming for all dimming lines, but this is not the case for all events. Generally, we prefer a larger magnitude dimming as its interpretation is less ambiguous and less susceptible to being dominated by other physical processes such as active region evolution. As was shown in \citet{Mason2014}, the ionization level is inversely proportional to depth of dimming. Thus, 171 \AA\ is generally preferred as the dimming line but is evaluated on a case by case basis for the events studied here. Finally, we prefer to use 284 \AA\ as the non-dimming line for deconvolution based on the physical motivation provided in the paragraph above.

In \citet{Mason2014}, it was found that for the 2010 August 7 event, the combination of Fe IX 171 \AA\ (dimming) and Fe XV 284 \AA\ (non-dimming) in EVE gave the best match to the spatially isolated dimming in AIA 171 \AA. The only dimming mechanisms identified to be important in this event were mass-loss and thermal. Thus, it seems that the 171 \AA\ - 284 \AA\ combination can successfully mitigate the impact of thermal processes on the dimming line. If other dimming mechanisms play an important role in the irradiance, as is the case for the 2011 August 4 case in \citet{Mason2014}, it may be necessary to account for them, such as by identifying and removing the impact of obscuration dimming. Until such an analysis is performed, we apply the deconvolution method to the additional 28 events with viable EVE data, using the clean removal of the flare peak as the criteria for determining the best combination of dimming-non-dimming line. In other words, the peaks of the dimming and scaled/time-shifted non-dimming lightcurves should be similar in shape. Figure \ref{fig:deconvolutioncombinations} shows that many of the combinations would meet this criteria. The next determining factor is depth of dimming. Event 20 had a relatively consistent depth of dimming for all dimming lines, but this is not the case for all events. Generally, we prefer a larger magnitude dimming as its interpretation is less ambiguous and less susceptible to being dominated by other physical processes such as active region evolution. As was shown in \citet{Mason2014}, the ionization level is inversely proportional to depth of dimming. Thus, 171 \AA\ is generally preferred as the dimming line but is evaluated on a case by case basis for the events studied here. Finally, we prefer to use 284 \AA\ as the non-dimming line for deconvolution based on the physical motivation provided in the paragraph above.

\section{Physics Motivation for Dimming-CME Correlations} 
\label{sec:physicsmotivation}

This section provides the mathematical derivations that link the physics to the expected observations, with assumptions made explicit. Section \ref{sec:depthmass} focuses on the relationship between dimming irradiance depth and CME mass and Section \ref{sec:slopespeed} on the relationship between dimming irradiance slope and and CME speed. 

\subsection{Dimming depth -- CME mass relationship}
\label{sec:depthmass}

\citet{Aschwanden2009a} provide a mathematical description of CME expansion and here we adapt it for the variation of intensity in collisionally-excited lines as a function of height for a constant-temperature and expanding volume. The emission measure is equal to the square of the density of the emitting material integrated over the emitting volume. The relative change as a function of time, $t$, in emission then becomes

\begin{equation}
\begin{split}
\label{eq:dimmingdensity}
    \frac{n_d(t)}{n_A} & = \sqrt{\frac{I_d(t)}{I_A}} \\ 
    \implies n_d(t) & = n_A \sqrt{\frac{I_d(t)}{I_A}}
\end{split}
\end{equation}

\noindent where $n_d(t)$ is the density in the dimming region as a function of time, $n_A$ is the density of the nearby active region responsible for triggering the dimming, and $I$ is the emission intensity.  The volume is assumed to be the void in the corona where the plasma existed before the CME stripped it away, so it falls out of the relative equation. Next, the corona consists primarily of hydrogen (at least 90\%) and at the known average temperature (about 1 MK), it can be reasonably assumed to be fully ionized. Because hydrogen consists of a single proton and single electron, the densities of these two populations, $n_p$ and $n_e$, are equal but the mass is dominated by the protons, $m_p$. Thus, the mass of the CME, $m_{CME}$, can be expressed as 

\begin{equation}
    m_{CME} = m_p n_e V_{CME}
\end{equation}

\noindent where $V_{CME}$ is the volume of the CME. The mass of the region in the corona that dims can be similarly expressed as 

\begin{equation}
\label{eq:dimmingmass}
    m_d(t, T) = m_p n_d(t) V_d
\end{equation}

\noindent where the subscript $d$ indicates the dimming and $m_d$ is a function of time and temperature, $T$. We can then substitute Equation \ref{eq:dimmingdensity} into Equation \ref{eq:dimmingmass} and obtain

\begin{equation}
    m_d(t, T) = m_p\left[n_A \sqrt{\frac{I_d(t)}{I_A}}\right] \frac{\pi}{4} w_d^2 \lambda_T
\end{equation}

\noindent where $w_d$ is the width of the dimming region and $\lambda_T$ is the hydrostatic scale height. Note that these equations as developed by \citet{Aschwanden2009a} are intended for use with EUV imagers. For EVE irradiance, we make the assumption that the temperature response is ``pure" as the lines of interest are spectrally resolved. This assumption comes with the caveat that, because EVE is spatially integrated, the temperature evolution of plasma outside of the dimming region can be ignored. Given the flare-dimming (i.e. temperature) deconvolution method described earlier and in \citet{Mason2014}, we believe this to be a reasonable assumption. Furthermore, later in this paper, we will be establishing a correlation between coronagraph-derived and dimming-derived masses. If the distribution of ions is assumed to be the same in each event, then it is not necessary for us to compute the dimming at each temperature since this will be captured in the empirical relationship between the two mass-estimation methods. We further assume the constancy of other terms as follows 

\begin{equation}
    k_A = m_p n_A \frac{\pi}{4} w_d^2 \lambda_T.
\end{equation}

\citet{Aschwanden2009a} makes the assumption that $w_d \approx w_A$, i.e., that the size of the dimming is about the same size as the active region. Active regions that produce large CMEs tend to be of similar size, varying by perhaps a factor of 2. Dimming regions tend to occur near the responsible active region and account for more area than the active region itself, so this should be considered a lower bound. The size of the dimming could of course be measured but our goal is to use irradiance dimming (i.e. spatially integrated) with no additional information and determine a relationship to CME kinetics for use in space weather applications. The assumption that the density of active regions is constant is difficult to assess. Thus, we can simplify the dimming mass equation to 

\begin{equation}
    m_d(t) = k_A \sqrt{\frac{I_d(t)}{I_A}}. 
\end{equation}

\noindent As mentioned before, this equation relates the absolute mass to the relative emission intensity. \citet{Aschwanden2009a} used the active region intensity for normalization. With EVE, we use the disk-integrated, pre-flare emission intensity (i.e., pre-flare irradiance), $I_{tot}$, for normalization. We also wish to parameterize the dimming in terms of a single number to relate to a single CME mass number to be determined with the traditional coronagraph-based methods. This will be herein referred to as dimming depth and is taken at a manually selected time. 

\begin{equation}
    m_d = k_A \sqrt{\frac{I_d(t=t_{selected})}{I_{tot}}}.
\end{equation}

Finally, we have an equation to estimate dimming mass from irradiance: 

\begin{equation}
    m_d = k_A \sqrt{depth}
\end{equation}

\noindent and we'll make the assumption that $m_d \approx m_{CME}$. 

\subsection{Dimming slope -- CME speed relationship}
\label{sec:slopespeed}
Again, we start with the emission measure equation but now consider the standard way to model gravitational stratification of the background corona using the multi-hydrostatic model in \citet{Aschwanden2004}. This provides an equation for the emission measure at a particular altitude, $h$, which is in the plane-of-sky and at a particular temperature. It is expressed in terms of the equivalent column depth, $z_{eq}$, 

\begin{equation}
    EM(h, T) = n_e^2(h_0, T) z_{eq}(h, T).
\end{equation}

\noindent Aschwanden goes on to model and compute $z_{eq}$. We make the simplifying assumption that $z_{eq} = h$ and consider the collisionally excited bound-bound emission for a self-similar, spherical expansion of plasma at constant temperature. Mass is conserved during the expansion: 

\begin{equation}
    n_0 h_0^3 = n(t) h^3(t)
    \Rightarrow n(t) = \frac{n_0 h_0^3}{h^3(t)}
\end{equation}

\noindent which can then be plugged into the emission measure equation and simplified to yield

\begin{equation}
    \frac{EM(t)}{EM_0} = \left(\frac{h_0}{h(t)}\right)^5. 
\end{equation}

\noindent Similarly to the starting equations in Section \ref{sec:depthmass}, these equations were originally developed for usage with EUV spectral imagers. Spatially integrating in the plane-of-sky reduces the exponent by 2, i.e., 

\begin{equation}
\label{eq:irradiancevsheight}
    \frac{I(t)}{I_0} = \left(\frac{h_0}{h(t)}\right)^3
\end{equation}

\noindent We then assume that when the irradiance has dropped halfway between its initial value and the dimming depth, the CME is at its average velocity. Note that acceleration must occur in the low corona. From the 25,000 (20 years worth) CMEs in the CDAW catalog, we know that above 2 R$_\odot$ (the approximate lower bound of the LASO C2 coronagraph), the average acceleration is -1.67 m s$^{-2}$ compared to average velocities of 393 km s$^{-1}$. So the acceleration of the CME from 0 to e.g., 393 km s$^{-1}$ must typically occur below 2 R$_\odot$. The acceleration mechanism is poorly understood and few observations exist that can be used to determine the CME velocity in the low corona. We assume no further acceleration at this point, i.e., 

\begin{equation}
    h(t) = h_0 + vt, 
\end{equation}
 
\noindent then substitute this into Equation \ref{eq:irradiancevsheight} and simplify to get 

\begin{equation}
\label{eq:velocitymidstep}
    v = - \frac{7 h_0}{8 t_{1/2}}
\end{equation}

\noindent $t_{1/2}$ can be obtained from the light curve of the event assuming the average slope as 

\begin{equation}
    t_{1/2} = \frac{1}{2}\left(\frac{I_0}{\frac{dI}{dt}}\right)
\end{equation}

\noindent where $\frac{dI}{dt}$ is the slope of the light curve while the dimming is in progress. Plugging this back into the velocity equation of Equation \ref{eq:velocitymidstep} and simplifying results in 

\begin{equation}
\begin{split}
      v & = - \frac{7}{4} h_0 \frac{\frac{dI}{dt}}{I_0} \\ 
    let\ H & = - \frac{14}{8} h_0  \\
    \Rightarrow v & = H \frac{\frac{dI}{dt}}{I_0}.
\end{split}
\end{equation}

\noindent $H$ is dependent on the initial size of the CME/dimming region. Herein, we assume that this value is constant but the value could be estimated with image data, just as $w_d$ could be for the mass-depth relationship of the previous section. Again however, we are interested in the development of a correlation between CMEs and dimmings where the dimmings are determined entirely by irradiance. $\frac{dI}{dt}$ is the slope and $I_0$ is the depth. Thus, the equation that we'll refer to in subsequent sections is

\begin{equation}
\label{eq:speedslope}
    v = H \frac{slope}{depth}. 
\end{equation}

\noindent Note that this method does not include any direction information, so the $v$ indicates speed rather than velocity. \citet{Aschwanden2009b} developed a more sophisticated model of dimmings, including adiabatic expansion and gravitational stratification. However, the model contains 14 free parameters and is more suited to a case-by-case study of dimming morphologies. For the purposes of our correlative study, it is reasonable to assume that the decrease in emission due to the volume density is more significant than the thermal and inhomogeneity effects, and that the effective height scale of the CME is the most important parameter. 

\section{Light Curve Fitting}
\label{sec:lightcurvefitting}
\begin{figure}
    \begin{center}
	    \includegraphics[width=90mm]{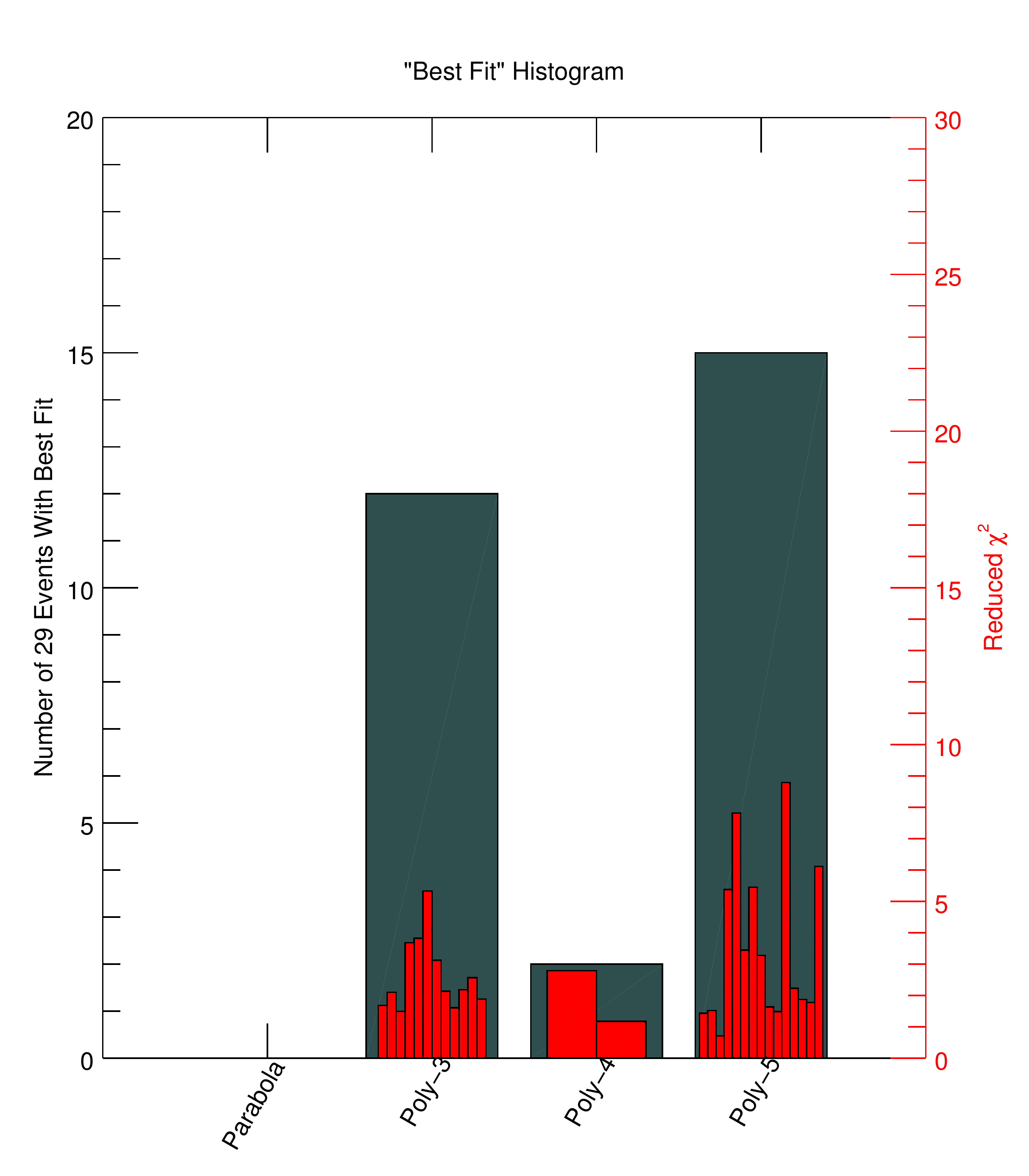}
    \end{center}
    \caption{
        (grey) Statistics of manually selected ``best fit" for all unique EVE dimming events in 4 weeks studied and 
        (red) the reduced $\chi^2$ for the best fits. The 3rd and 5th order polynomial fits provided the largest number
        of best fits.
    }
    \label{fig:bestfithistogram}
\end{figure}

\begin{figure}
    \begin{center}
	    \includegraphics[width=90mm]{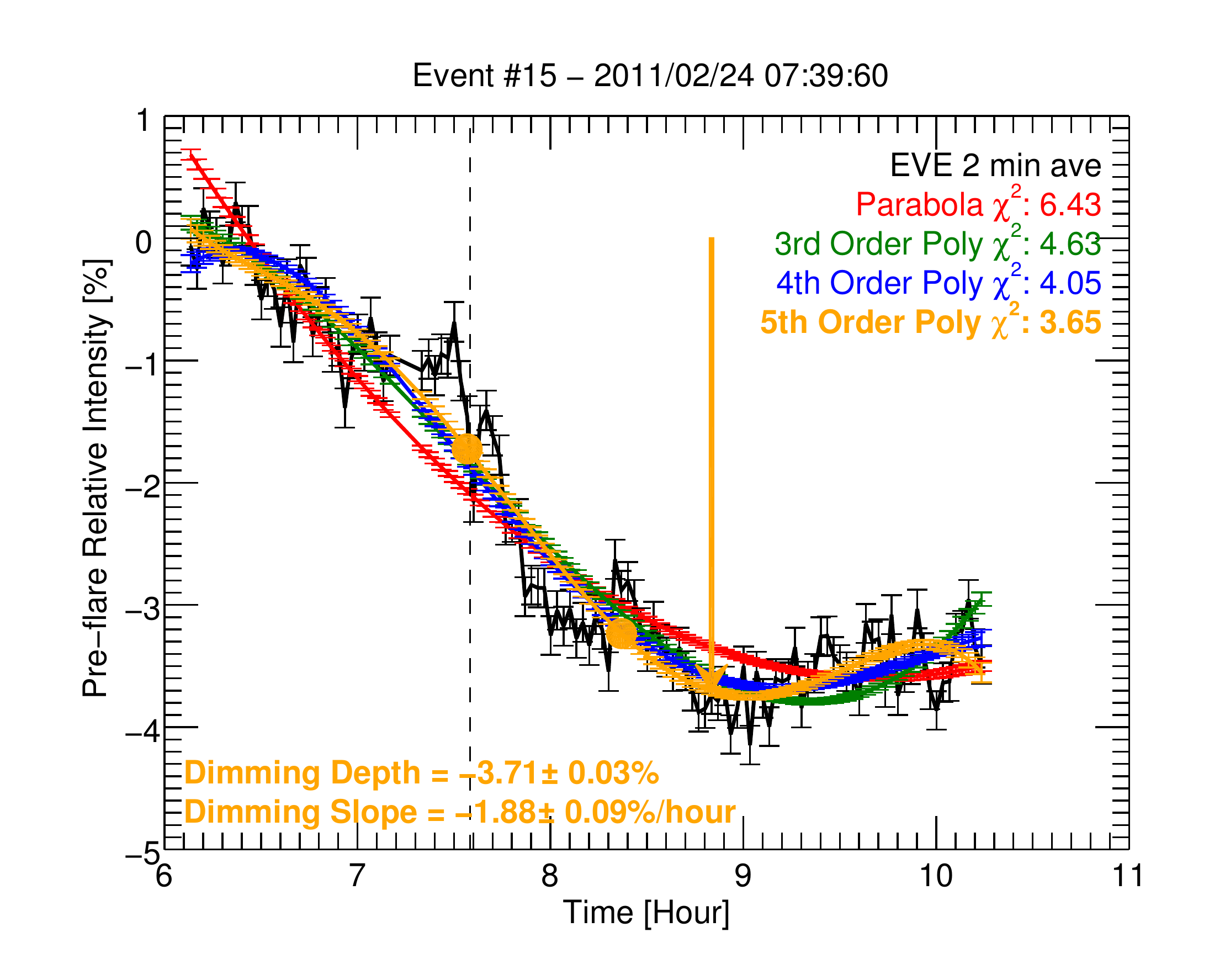}
    \end{center}
    \caption{
        A single dimming event (Event 15) showing the reduction in uncertainties of the fits compared to the EVE data. The
        arrow shows the location of dimming ``depth" parameterization for this event, and the two filled circles indicate the 
        range where ``slope" was computed. Their colors correspond to the fit types in the legend. The lowest $\chi^2$ indicates 
        that the 5th order polynomial was the best fit for this event, but we note that the results from the other 
        polynomial fits are very similar.
   	}
    \label{fig:fitsexample}
\end{figure}

Different functions were fitted to the EVE dimming events to explore which functions are more optimal for determining the dimming event parameters of depth and slope. Exponential and power law fits tend to result in $\chi^2 > 20$, meaning they were very poor fits. Polynomial fits up to order five were also computed, with 5th and 3rd orders appearing to best describe the shape of the light curves (see Figures \ref{fig:bestfithistogram} and \ref{fig:fitsexample}). The manually-selected best-fit function for each event was used for deriving the dimming slope and depth (see Section \ref{sec:parameterization}). 

\subsection{Dimming Fit Uncertainty Computation}

Coronal dimming is a transient event lasting several hours that is studied in terms of relative change from the initiation time. As such, no long-term degradation of SDO/EVE needs to be factored into uncertainties, i.e., the absolute accuracy is not important but the measurement precision is most important for the dimming uncertainty. To estimate precision, a period of solar inactivity was analyzed: 2013 January 28 from 00:00 -- 01:00 UT. The estimated precision of these 120-sec averaged SDO/EVE line data was calculated as the variance of the mean, i.e., the standard deviation divided by the square root of the number of samples, which was 12 in this analysis \citep{Bevington2003}. Table 2 provides the estimated precision for each emission line used in this study, and provides a sense of how well we can detect SDO/EVE dimmings that have depths less than 5\% of the pre-flare irradiance level. 

These base uncertainties were propagated through each step of the SDO/EVE dimming correction method described in \citet{Mason2014}, which were finally fed as measurement errors into IDL's \textit{poly{\_}fit} function for fitting the dimming trend. Figure 3 shows the comparison of the measurement errors and the resultant 1$\sigma$ uncertainties computed by \textit{poly{\_}fit}. The fits achieve the desired effect of reducing uncertainty even further than 120-sec averaging of the EVE data and providing a smooth function to parameterize. In particular, the fits smooth out any residual bumps in the light curve that the temperature-evolution correction method did not remove or that it introduced.

\begin{deluxetable}{ccc}
\label{tab:evelineprecision}
\tablecaption{Estimated precision for selected emission lines in EVE spectra. The Fe IX 171 \AA\ and Fe XV 284 \AA\ emission lines
              are the choice lines for dimming analysis with EVE data.}
\tablecolumns{3}
\tablewidth{0pt}
\tablehead{
\colhead{Ion} & \colhead{Wavelength [\AA]} & \colhead{Estimated} \\
\colhead{}    & \colhead{}                 & \colhead{Precision [\%]}
}
\startdata
	Fe IX & 171 & 0.25 \\ 
	Fe X & 177 & 0.21  \\ 
	Fe XI & 180 & 0.16 \\ 
	Fe XII & 195 & 0.20 \\ 
	Fe XIII & 202 & 0.21 \\ 
	Fe XIV & 211 & 0.32 \\ 
	Fe XV & 284 & 0.35 \\ 
	Fe XVI & 335 & 0.86 \\ 
	Fe XVIII & 94 & 0.42 \\
	Fe XX & 132 & 1.00 
\enddata
\end{deluxetable}

\section{Dimming and CME Parameters Correlation}
\label{sec:parameterization}

As described in Section \ref{sec:physicsmotivation}, we expect CME mass to be directly proportional to the square root of dimming depth, and dimming slope to be related to CME speed. While our intention for this study was to have ~30 events to test these hypotheses, it was challenging to obtain CME speed and mass for all of the candidate events. This section first describes the methods for determining the dimming and CME parameters and then presents the results from comparing the two sets of parameters. 

\subsection{Method for Deriving Dimming and CME Parameters}
The dimming parameters of depth and slope are determined by using manually selected best fits to SDO/EVE dimming light curves as described in Section \ref{sec:lightcurvefitting}. This approach also manually selected points for the computation of slope and depth. The former was guided by the desire to have $\chi^2$ near unity and by some flexibility for events where the SDO/EVE dimming correction method did not completely remove the flare peak of the cool corona line (Fe IX 171 \AA). In such cases with a residual flare peak, the fits can deviate from the ``pure" dimming light curve and skew the   upward. Rather than develop a complicated algorithm to account for this effect autonomously, selection of the best fit was done by manual inspection. Dimming slope was computed across a range: the initial point was typically chosen to be soon after the initial dimming rollover when the slope became relatively constant, and the final point was selected just prior to the inverse rollover leading to the relatively flat period in the light-curve (see solid circles Figure \ref{fig:fitsexample}). The slope is not necessarily constant between these two points. For each time step within the selected range, the derivative was computed. The single-value slope parameter for each event is the mean of these derivative (slope) values. The dimming depth parameter is taken from a relatively stable pre-flare value to a point near the beginning of the dimming floor (see arrow in Figure \ref{fig:fitsexample}). 

The detailed 3-D analysis of the speed and mass was possible for six of the best-observed CMEs, using combinations of the three coronagraphs (SOHO/LASCO, STEREO-A/COR, and STEREO-B/COR). These six events are shown as solid red symbols in Figure \ref{fig:correlations}. Following the method of \citet{Colaninno2009}, the GCS model is fit to the observations to determine the 3-D location and heights of the CMEs. The 3-D heights and longitude of the CME are needed to calculate the ``true" 3-D mass of the CME. These heights are also used to calculate the de-projected velocity of the CME. The reported masses are for a height of 15 R$_\odot$, using the fitting method of \citet{Bein2013} for mass increase with height. For the 2011 February 13-15 CMEs the mass was measured in both STEREO-A/COR2 and STEREO-B/COR2 and then averaged. For the 2011 August 9 and 11 CMEs, the mass was measured in LASCO/C3 only.

The following procedure was used to estimate the uncertainties for the CME kinetic parameters. The LASCO CDAW measurements were used for most of the events to derive the CME speed and mass, which are based on a single viewpoint observation as opposed to 3-D. The reported linear speed of each CME is obtained by fitting a straight line to the height-time measurements at a fixed position angle. If we assume that the CME axis is 60\degr\ from the sky plane as the worst case (for non-halo CMEs), this results in a factor of 2 (50\%) underestimation of the speed. The CDAW catalog also provides the CME span angle, which can be used to provide an estimated error on the CME mass (Figure 4 of \citet{Vourlidas2010a}. As an example, if we take Event 2 on our list, then using these errors we have 338 $\leqslant$ velocity $\leqslant$ 345 km s$^{-1}$ and 3.40 x 10$^{14}$ $\leqslant$ mass $\leqslant$ 4.30 x 10$^{14}$ g. 

For the six events with 3-D analysis of the CME measurements, we derive the error in the speed from the linear fit to the data assuming the error in the 3-D height measurements is $\pm$0.48 R$_\odot$ \citep{Colaninno2013}. Thus, if we take Event 7 as a typical 3-D CME measurement, we get 353 $\pm$ 13 km s$^{-1}$ for the speed. The mass is still considered an underestimate from the 3-D analysis but is better determined because the plane-of-sky angle and 3-D heights are known from the GCS model fit, so a $\pm$15\% error is assumed for the 3-D mass estimates \citep{Bein2013}.

For the purpose of correlating with dimming parameters, the midpoint between the low and high limits is chosen for each CME speed and mass parameter reported here, and the CME parameter error is the range between the high and low limits divided by two (i.e., $\pm$ error bars in Figure \ref{fig:correlations}). The plot of the points themselves does not display this center-point for single-viewpoint derived CME parameters but does for 3-D derived CME parameters. 

\subsection{Comparison of Dimming-CME Parameters}
\label{sec:comparison}

As described in \citet{Mason2014}, we expect direct proportionality between dimming depth and CME mass, and between dimming slope and CME speed. This relationship is intuitive, but Section \ref{sec:physicsmotivation} derived the functional form of the relationship. Namely, we expect that CME speed goes as dimming slope/depth and CME mass goes as the square root of depth. In other words, there should be a stronger correlation between these parameters than between any other combination of parameters. Table \ref{tab:correlations} provides the Pearson correlation coefficients \citep{Pearson1895} and p-value permutation statistical tests between each permutation of the dimming and CME parameters. Smaller p-values indicate a lower probability that the correlation could have arisen if no correlation existed at all. There is positive correlation between all of the parameter permutations, which is likely due to ``big flare syndrome \citep{Kahler1982, Kahler1992}, e.g., a rapid, powerful coronal magnetic field energy release tends to result in a faster, more massive CME.

\begin{deluxetable}{cccc}
\label{tab:correlations}
\tablecaption{Pearson correlation coefficients (PCC) and p-values between dimming and CME parameters.}
\tablecolumns{4}
\tablewidth{0pt}
\tablehead{
\colhead{Parameter 1} & \colhead{Parameter 2} & \colhead{PCC} & \colhead{p-value}
}
\startdata
	Slope/Depth & Speed & 0.12 & 0.65 \\ 
    $\rm \sqrt{Depth}$ & Mass & 0.75 & $4.30 \times 10^{-4}$ \\
	Slope & Speed & 0.78 & $1.51 \times 10^{-4}$ \\
	Depth & Mass & 0.74 & $7.80 \times 10^{-4}$ \\
	Slope & Mass & 0.60 & 0.01 \\
	Depth & Speed & 0.51 & 0.04 \\
	Mass & Speed & 0.64 & $2.79 \times 10^{-3}$ \\
	Slope & Depth & 0.27 & 0.15
\enddata
\end{deluxetable}

Our expectation was that we would have the highest correlations between the parameters that had a physical motivation for existing (see the first two rows of Table \ref{tab:correlations}). This was accurate for the $\rm \sqrt{depth}$ -- mass relationship, which had the second highest correlation of any two parameters at 0.75. The relationship was effectively just as strong directly between depth and mass, at 0.74. However, the slope/depth -- speed relationship performed worse than any other parameter combination, at 0.12. This is likely due to the numerous assumptions that were made during the derivation of that relationship. Those assumptions were made explicit in the derivation and they can be tested in future work. Interestingly, the best performing correlation was between slope and speed. Here, the implicit assumption is that the functional relationship between the two is a simple linear one. The direct proportionality between the two is intuitive, but that proportionality doesn't indicate whether speed should go as, e.g., the square of the slope. Our derivation based on the physics of the corona was supposed to provide that functional form, but it appears the assumptions made negatively impacted the correlation. The assumption of the initial dimming size, $h_0$, being about the same for all events is the primary suspect for why the derived CME speed relationship to slope/depth is not well correlated. Another way of assessing the correlation is through scatterplots and linear fits.

\begin{figure*}
    \begin{center}
	    \includegraphics[width=\textwidth]{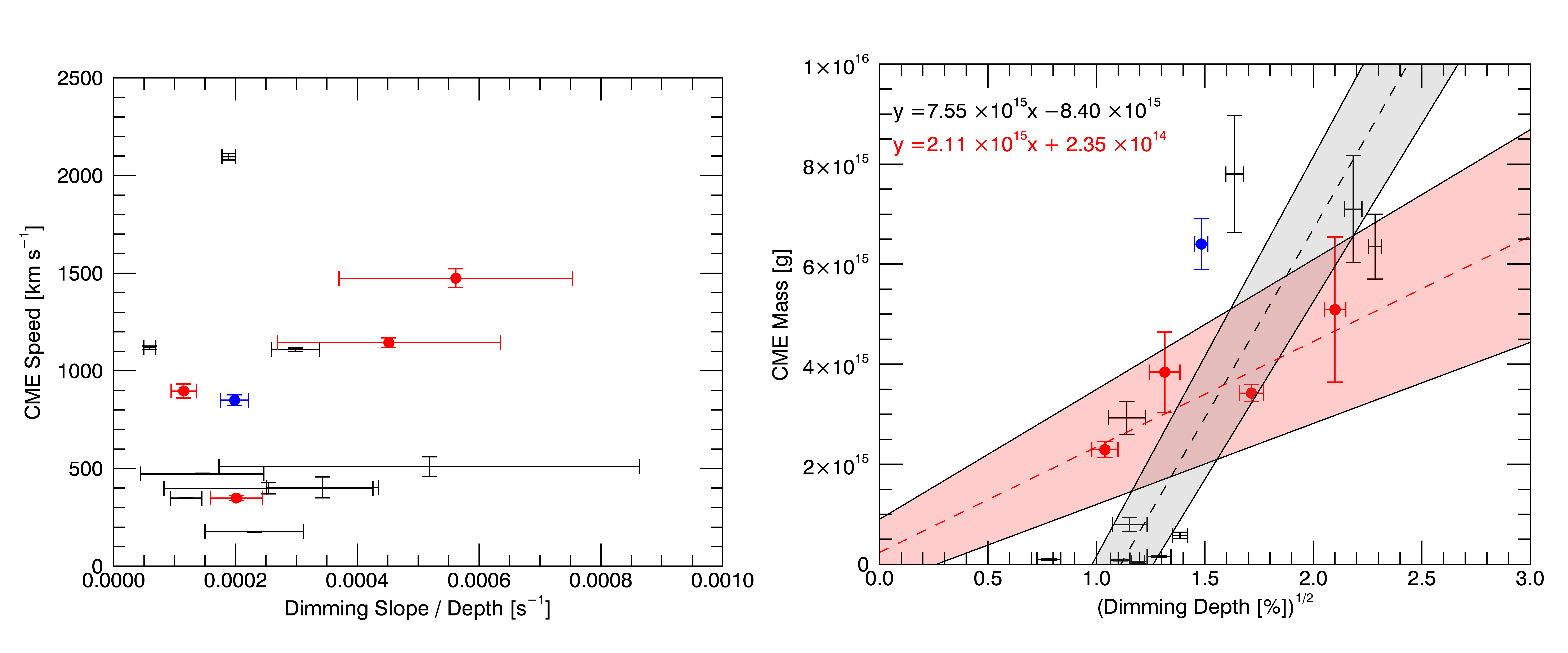}
    \end{center}
    \caption{
        Scatterplots of (left) CME speed and dimming slope/depth and (right) CME mass and dimming $\rm \sqrt{depth}$. 
        Data without a 
        center-point are derived from a single viewpoint of CMEs and are thus presented as a range of possible values 
        rather than a single point with a standard uncertainty. Red symbols, line, and text indicate 3-D computed CME 
        parameters, and the blue symbol indicates data from the simple 2010 August 7 event, which is also 3-D derived. 
        Linear fits are shown as the dashed lines, and the grey/pink region represents the $1\sigma$ uncertainty of 
        the linear fits.
   	}
    \label{fig:correlations}
\end{figure*}

Figure \ref{fig:correlations} shows scatterplots of speed vs. slope/depth and mass vs. $\rm \sqrt{depth}$ with estimated error bars. Linear fits for the latter were computed using IDL's \textit{fitexy}, which can accept input errors in both axes and return the fit parameters with a $1\sigma$ uncertainty. The fit uncertainty is then used to define the grey/pink regions of Figure \ref{fig:correlations}. The fit equations are also listed in the Figure \ref{fig:correlations} panels. This process was repeated using only CME values computed from the 3-D methods and are plotted as the red dashed line and pink shaded region. In order to get a nominal fit for the 3-D case with so few data points, a virtual (0, 0) point was added to the fit. The same fitting procedures were also applied to the speed vs. slope/depth plot, but the fits were extremely poor as expected based on the low Pearson correlation coefficient and inspection of the scatter. 

\begin{figure*}
    \begin{center}
	    \includegraphics[width=\textwidth]{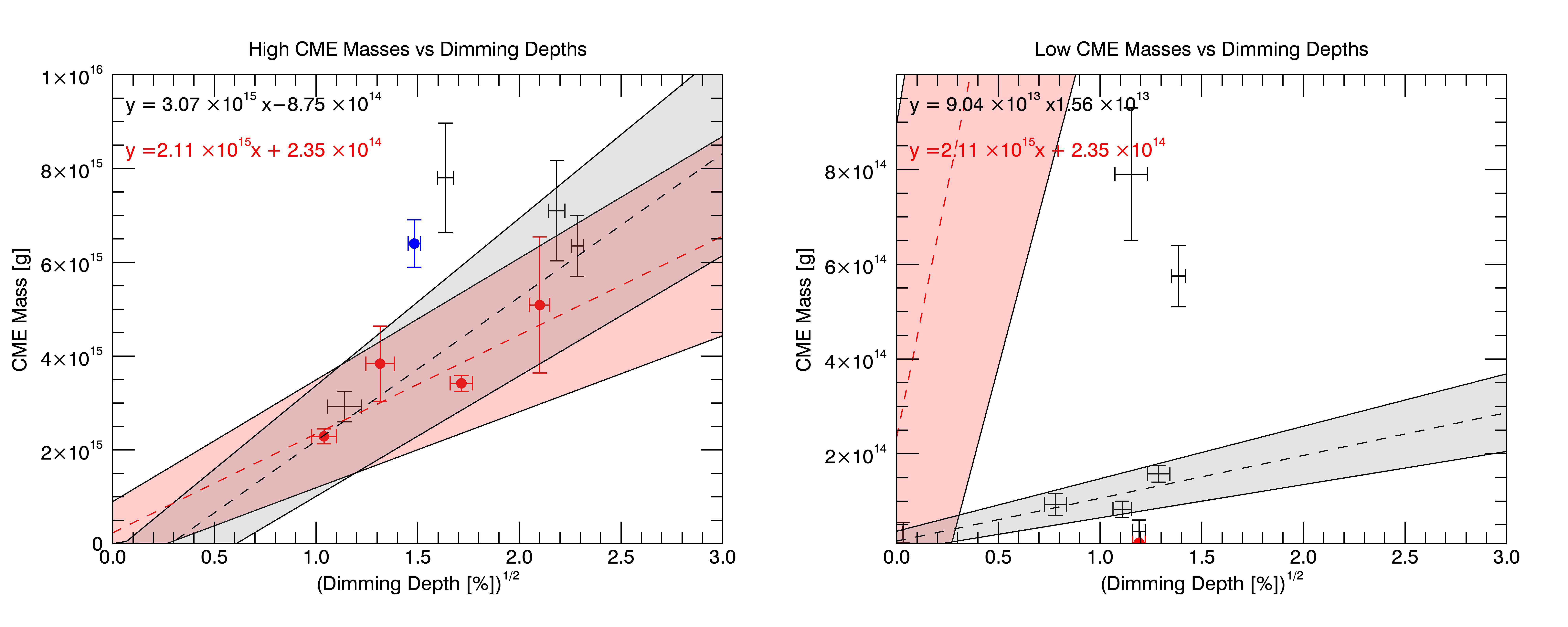}
    \end{center}
    \caption{
        Same as Figure \ref{fig:correlations} but for (left) high CME mass ($\geq\ 1 \times 10^{15}\ g$) (right) low CME 
        mass ($< 1 \times 10^{15} g$). 
   	}
    \label{fig:correlationshighlowmass}
\end{figure*}

The mass vs. $\rm \sqrt{depth}$ plot (Figure \ref{fig:correlations}, right) is linear-linear for clarity of the fits, but several of the data points end up off scale as they are $< 1 \times 10^{15}\ g$. These points skew the fit significantly. Figure \ref{fig:correlationshighlowmass} shows the fit applied to high-mass only and low-mass only separately, with the 3-D based fit from Figure \ref{fig:correlations} still shown in red. The high-mass only plot shows very good agreement between the fits for all points and 3-D points, with slopes agreement to 37\% of each other, whereas the fit for low-mass only has a fit slope that is 2 orders of magnitude lower than both the 3-D fit and the high-mass fit. Thus, we suspect there may be two statistical families in the data. We examined all of these events individually but did not notice any dimming peculiarities that might cause this separation of high-mass and low-mass families in this comparison. We also verified that the families do not strongly correlate to GOES flare magnitude (or whether there was a flare at all), CME span, or flare type. There may be a systematic error in the mass-estimation method that becomes acute in conditions that result in low masses. Furthermore, the low mass CMEs seem to be out of family when compared to an independently derived relationship between flare energy and CME energy established by \citet{Emslie2012}.

\begin{figure}
    \plotone{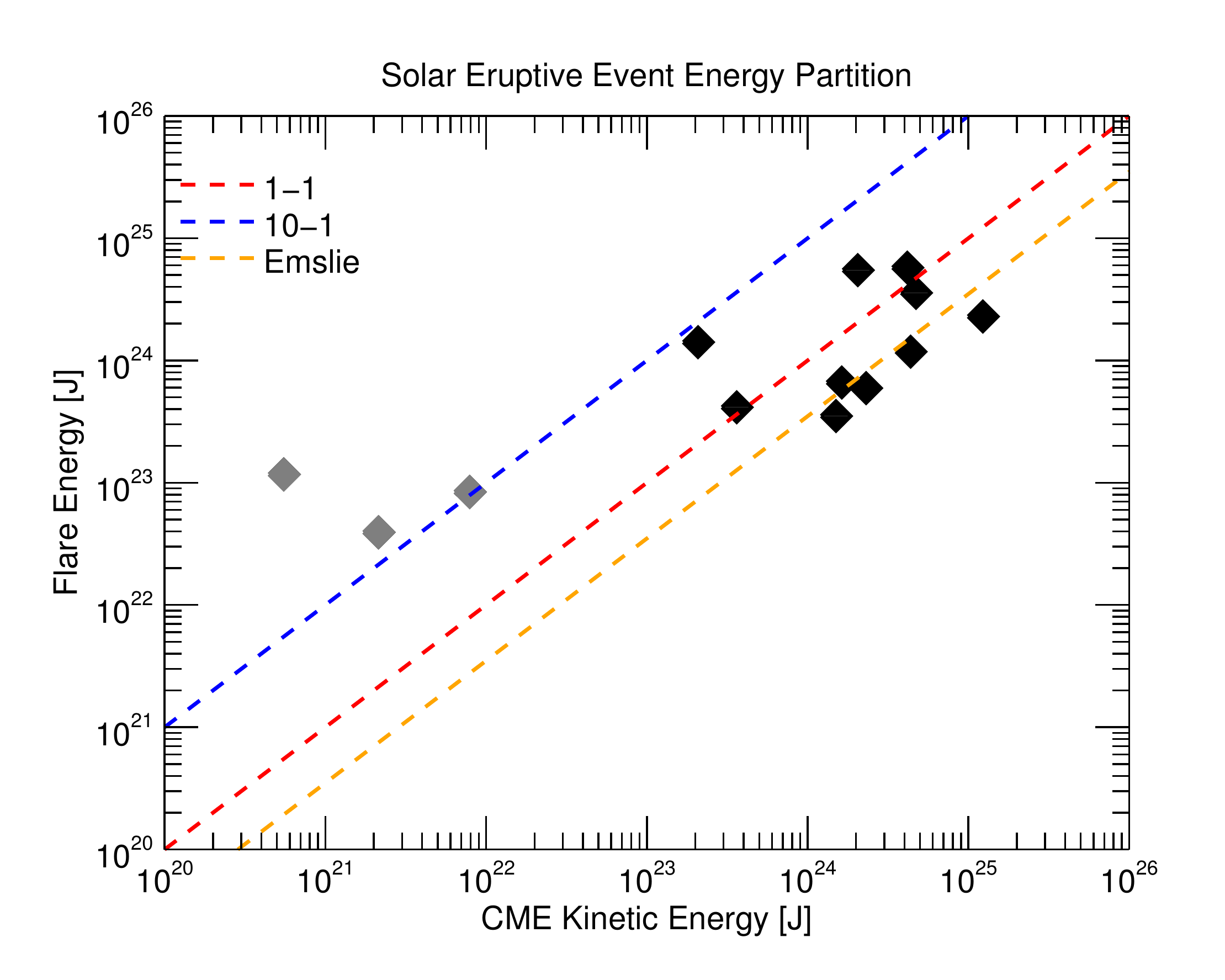}
    \caption{
        Flare-CME energy partition for the events of the semi-statistical study. The grey points are of the ``low-mass" 
        family of Figure \ref{fig:correlationshighlowmass} and the black points are from the ``high-mass" family. 
        The expectation is that the points should align roughly with the Emslie (0.35-1) line.  
    }
    \label{fig:flarecmeenergypartition}
\end{figure}

Figure \ref{fig:flarecmeenergypartition} shows a scatterplot of estimated flare energy versus CME kinetic energy. Flare energy was computed using the method in \citet{Woods2006}, which integrates the GOES XRS-B light curve over the period of the flare, multiplies that value by a 1 AU scaling factor and an additional empirically-determined scaling factor. Most, but not all, of the events studied here had associated flares. The coronagraph-determined CME mass and speed were combined to compute CME kinetic energy using the $KE = \frac{1}{2}mv^2$ equation. Figure \ref{fig:flarecmeenergypartition} also shows the expectation for the scatter-points based on the results of \citet{Emslie2012}, who determined that their 38 flare-CME events roughly fell between a 1-1 line and a 0.35-1 line. Our coronagraph-based high-mass results are consistent with \citet{Emslie2012} but the low-mass CMEs are 1-2 orders of magnitude lower than expected. This comparison suggests that the low-mass CME values may just be representative of lower limits for mass. Indeed, this is the qualifier provided by the traditional CME mass estimation community. 

\begin{figure}
    \plotone{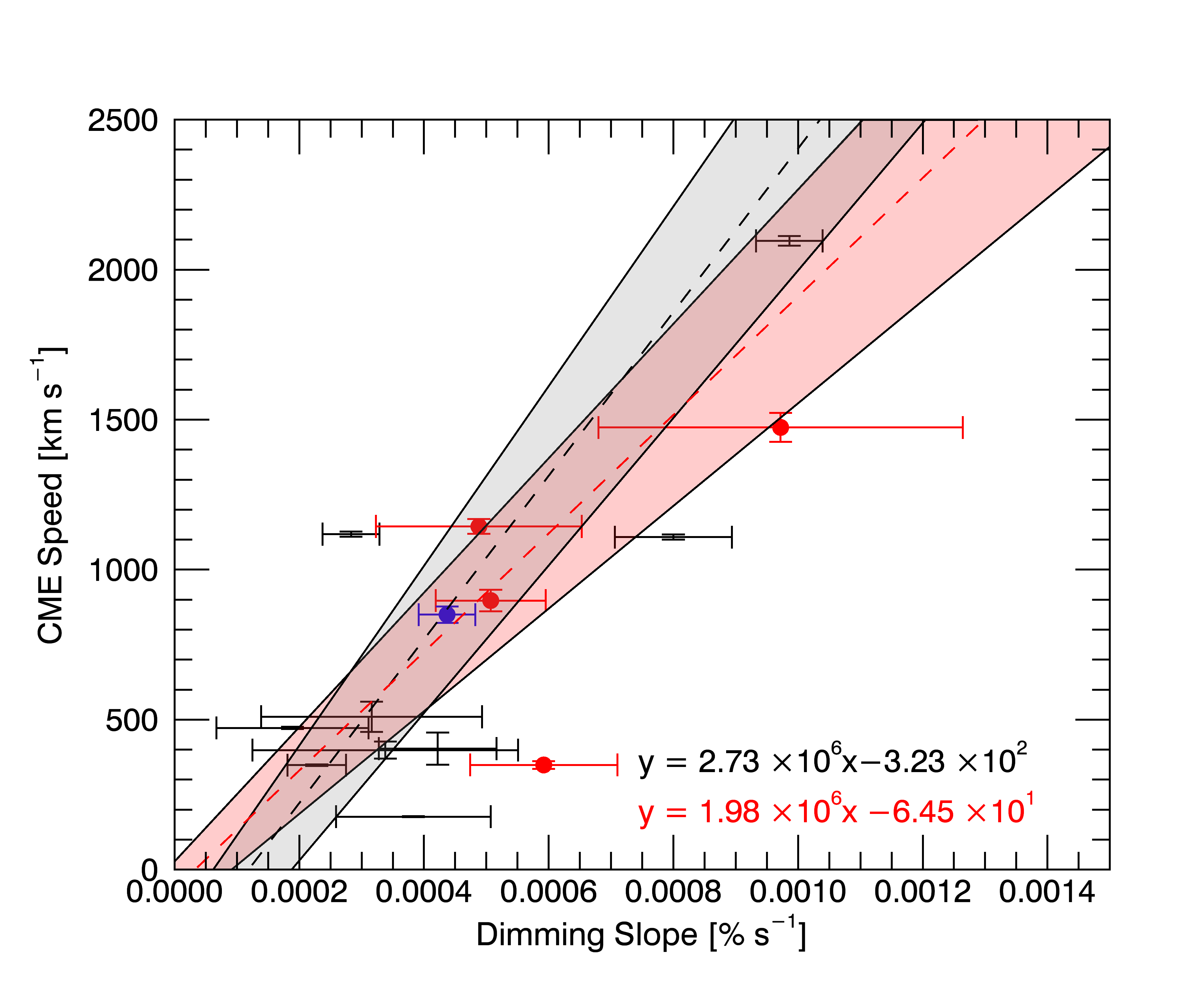}
    \caption{
        Same as Figure \ref{fig:correlations} (right) but the horizontal axis is simply slope rather than slope/depth. 
    }
    \label{fig:speedvsslope}
\end{figure}

We can also compare estimated CME energy through these two independent methods. The flare energies computed for our events (vertical axis of Figure \ref{fig:flarecmeenergypartition}) can be converted into estimated CME total energy using the relationship from \citet{Emslie2012}. The CME kinetic energy can also be estimated by using the relationships between dimming and CMEs established here. The equations of fit in Figure \ref{fig:correlationshighlowmass} for the high-mass and 3-D mass can be averaged to obtain an estimated CME mass based on dimming of 

\begin{equation}
\label{eq:massdepth}
    m_{CME} = 2.59 \times 10^{15} \sqrt{depth}
\end{equation}

\noindent where $m_{CME}$ is in units of g and depth is in relative units of \%. Note that the 3-D points are weighted more heavily since they show up in the high-mass as well as the 3-D populations. This is desired because the 3-D derived CME parameters are more trustworthy. We've also dropped the y-intercept, making the assumption that it should be 0 (the average is $-3.2 \times 10^{14}$, but this value is so small as to be lost in the noise anyway). In order to calculate the kinetic energy of the CME, we also need the estimated speed based on dimming. As described earlier, the physically-motivated relationship did not pan out but the Pearson correlation coefficient between CME speed and dimming slope was the highest of any parameter combination. Figure \ref{fig:speedvsslope} shows a scatterplot of this relationship in the same form as Figure \ref{fig:correlations}. 

\begin{figure}
    \plotone{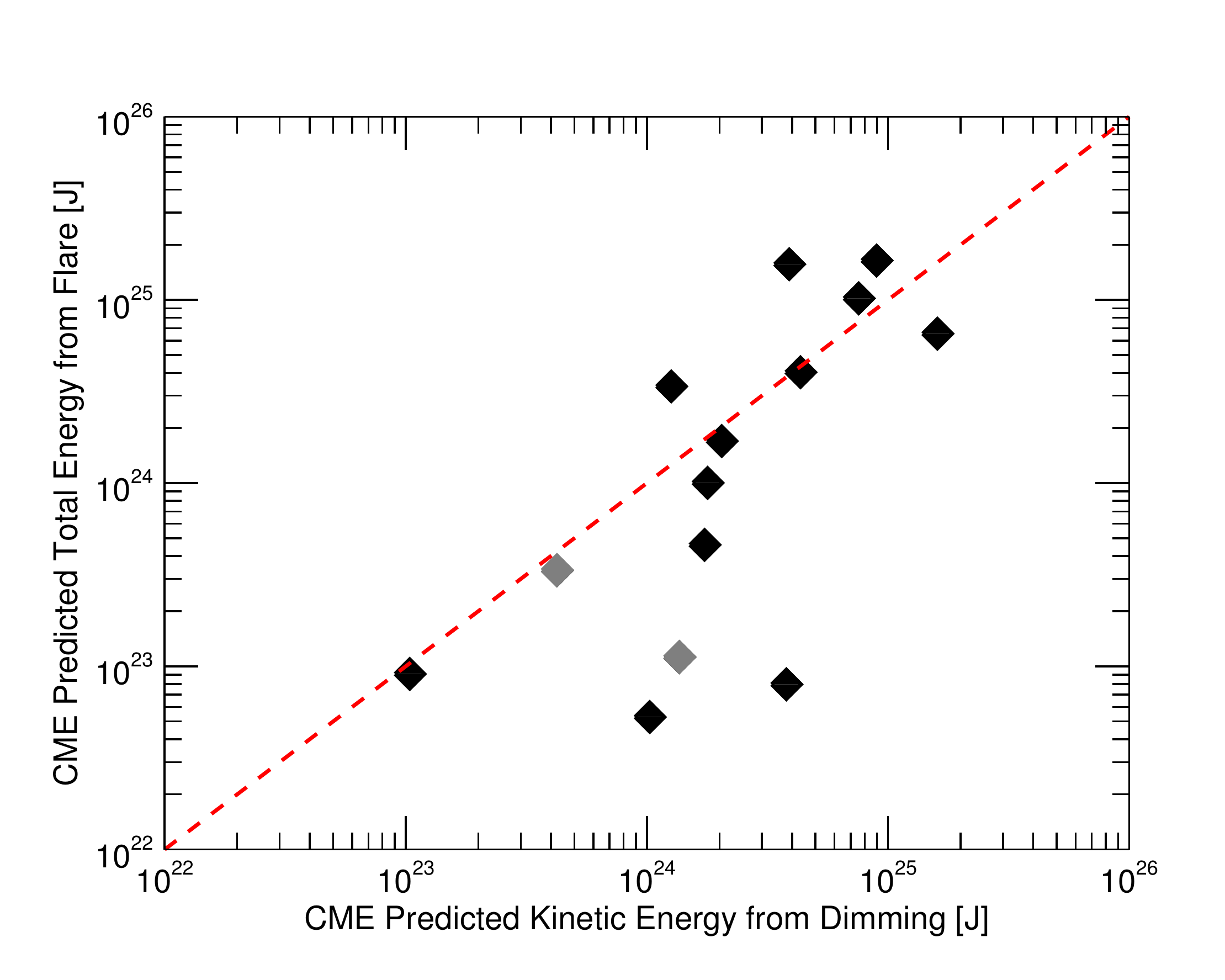}
    \caption{
        Scatterplot comparing CME energy estimation methods. The vertical axis uses the \citet{Emslie2012} result to 
        estimate CME total energy based on total flare energy, which was itself computed using the \citet{Woods2006} 
        method. The horizontal access is also estimated CME energy but based on the relationship established here 
        between dimming depth and CME mass (excluding the low-mass family). The red diagonal line indicates a 1-1 
        relationship. The grey points are ones that had low CME masses in the coronagraph-based mass estimation. 
    }
    \label{fig:cmepredictedvscomputed}
\end{figure}

The points here are significantly more linear, though still not perfect. More points are needed to gain greater statistical significance and determine if this relationship is real, but that is future work. For now, we will take the average of the fit to all points and the fit to 3-D points to obtain

\begin{equation}
\label{eq:speedslope}
    v_{CME} = 2.36 \times 10^6 slope
\end{equation}

\noindent where $v_{CME}$ is in units of km s$^{-1}$ and slope is in units of \% s$^{-1}$. The 3-D points are again weighted more heavily and the y-intercept is again dropped. Using Equations \ref{eq:massdepth} and \ref{eq:speedslope}, we can compute the kinetic energy of the CME using the $KE = \frac{1}{2}mv^2$ equation again. Combining this with the CME energy estimation based on flare energy discussed above results in Figure \ref{fig:cmepredictedvscomputed}. The Emslie/Woods method (vertical axis) determines the total CME energy i.e. kinetic + potential energy. The dimming-based CME-energy estimation only provides CME kinetic energy. However, gravitational potential energy is comparatively negligible, being only 9\% of the kinetic energy on average for the 38 events in \citet{Emslie2012}. Most of the points, except for three, cluster around the 1-1 line (red-dashed line). As this comparison relies only on the high-mass CME family, it suggests that the low-mass CME values are not realistic values but instead may represent the lowest limit of the estimation method. 

Note that uncertainties are not factored into the Pearson correlation coefficients quoted in Table \ref{tab:correlations}. Future work could use additional techniques for correlation that account for uncertainty, e.g., rank order. Such a study could include many more events to maximize the efficacy of the correlation comparison.

\section{Summary and Conclusions}
Positive correlations with a high degree of significance have been found between coronal dimming and CME parameters. Our physically-motivated hypothesis that the CME mass goes as $\rm \sqrt{depth}$ had the second highest correlation and the scatterplots looked good when coronagraph-based masses below $10^{15}$ g were ignored. The second hypothesis, that CME speed should go as dimming slope/depth, was proven incorrect (barring a very unlikely sampling of the statistical space). However, the direct relationship between CME speed and dimming slope had a strong Pearson correlation coefficient and strong significance, though the scatterplot showed that there is a need for more data points. Future work will include hundreds to thousands of events, which should alleviate any concerns about statistical significance. Nevertheless, tentative equations relating CME mass and speed to EUV irradiance dimming depth and slope have been established in Equations \ref{eq:massdepth} and \ref{eq:speedslope}. Additionally, we found that the Fe IX 171 \AA\ dimming corrected for the flare contributions using the Fe XV 284 \AA\ line provides the most accurate dimming results for the SDO/EVE data. We note that the uncertainties for coronagraph and dimming parameters are complimentary: there are smaller uncertainties for CME speed than dimming slope, and there are smaller uncertainties for dimming depth than CME mass.

\acknowledgments
The authors would like to thank Jim Klimchuk for discussions about the physical motivation for mass-loss dimming and Amir Caspi for identifying the need for a new mathematical derivation to establish the expected relationships between dimming and CME parameters. The CDAW CME catalog is generated and maintained at the CDAW Data Center by NASA and The Catholic University of America in cooperation with the Naval Research Laboratory. SOHO is a project of international cooperation between ESA and NASA. This research is supported by the NASA SDO project and NASA grant NAS5-02140. Author D. F. Webb was supported by Navy grant N00173-14-1-G014. Authors R. C. Colaninno and A. Vourlidas were supported by NASA contract S-136361-Y to the Naval Research Laboratory, and CNR funds.

\software{IDL}
\bibliography{./library}{} 
\bibliographystyle{./aasjournal}

\end{document}